\documentclass[sigconf]{acmart}

\AtBeginDocument{%
 }

\copyrightyear{2024}
\acmYear{2024}
\setcopyright{rightsretained}
\acmConference[CIKM '24]{Proceedings of the 33rd ACM International Conference on Information and Knowledge Management}{October 21--25, 2024}{Boise, ID, USA}
\acmBooktitle{Proceedings of the 33rd ACM International Conference on Information and Knowledge Management (CIKM '24), October 21--25, 2024, Boise, ID, USA}
\acmDOI{10.1145/3627673.3680008}
\acmISBN{979-8-4007-0436-9/24/10}

\makeatletter
\gdef\@copyrightpermission{
  \begin{minipage}{0.3\columnwidth}
  \end{minipage}\hfill
  \begin{minipage}{0.7\columnwidth}
   \href{https://creativecommons.org/licenses/by-nc-sa/4.0/}{This work is licensed under a Creative Commons Attribution-NonCommercial-ShareAlike International 4.0 License.}
  \end{minipage}
  \vspace{5pt}
}
\makeatother

\usepackage{tabularx}
\usepackage{amsfonts}
\usepackage{subcaption}
\usepackage{orcidlink}

\newtheorem{problem}{Problem}
\newtheorem{task}{Task}

\begin{document}

\title{\textsc{DeepClair}: Utilizing Market Forecasts for Effective Portfolio Selection}






\author{Donghee Choi}
\authornote{These authors contributed equally to this work.}
\orcid{0000-0002-8857-9680}
\affiliation{%
  \institution{Imperial College London}
  \department{Department of Metabolism, Digestion and Reproduction}
  \city{London}
  \country{United Kingdom}
}
\email{donghee.choi@imperial.ac.uk}

\author{Jinkyu Kim}
\authornotemark[1]
\orcid{0000-0001-9484-7541}
\affiliation{%
  \institution{Korea University}
  \department{Department of Computer Science}
  \city{Seoul}
  \country{South Korea}
}
\email{no100kill@korea.ac.kr}

\author{Mogan Gim}
\authornote{This work was done while the author was a postdoctoral researcher at Korea University.}
\orcid{0000-0002-6458-7723}
\affiliation{%
  \institution{Hankuk University of Foreign Studies}
  \department{Department of Biomedical Engineering}
  \city{Yongin}
  \country{South Korea}
}
\email{gimmogan@hufs.ac.kr}

\author{Jinho Lee}
\orcid{0000-0003-3076-3722}
\affiliation{%
  \institution{Shinhan Bank}
  \city{Seoul}
  \country{South Korea}
}
\email{jinholee@korea.ac.kr}

\author{Jaewoo Kang}
\authornote{Corresponding author}
\orcid{0000-0001-6798-9106}
\affiliation{%
  \institution{Korea University}
  \department{Department of Computer Science}
  \city{Seoul}
  \country{South Korea}
}
\email{kangj@korea.ac.kr}

\renewcommand{\shortauthors}{Donghee Choi, Jinkyu Kim, Mogan Gim, Jinho Lee, \& Jaewoo Kang}

\begin{abstract}

Utilizing market forecasts is pivotal in optimizing portfolio selection strategies. We introduce \textsc{DeepClair}, a novel framework for portfolio selection. \textsc{DeepClair} leverages a transformer-based time-series forecasting model to predict market trends, facilitating more informed and adaptable portfolio decisions. To integrate the forecasting model into a deep reinforcement learning-driven portfolio selection framework, we introduced a two-step strategy: first, pre-training the time-series model on market data, followed by fine-tuning the portfolio selection architecture using this model. Additionally, we investigated the optimization technique, Low-Rank Adaptation (LoRA), to enhance the pre-trained forecasting model for fine-tuning in investment scenarios. 
This work bridges market forecasting and portfolio selection, facilitating the advancement of investment strategies.

\end{abstract}




\maketitle


\section{Introduction}
Market outlooks are beneficial for achieving successful investment returns in portfolio selection. 
Classic portfolio selection strategies involve not only analyzing current market situations but also utilizing future market states~\cite{markowitz1959portfolio,lai1991portfolio,konno1991mean,rockafellar2000optimization}. 
The role of market forecasts in portfolio selection is renowned for its reliability in various sources from both academic communities and business industries~\cite{corberan2023portfolio_forecasting_genetic,wang2020portfolio_forecasting,yan2020portfolio_forecasting_lstm}.
Specifically, market forecasts can contribute to helping investors understand macroeconomic trends, diversify their assets, mitigate risks, and ultimately achieve better returns~\cite{tanaka1999portfolio,liu2006survey_credibility,corberan2023portfolio_forecasting_genetic}.

Previous research works have proposed portfolio selection frameworks that employ deep reinforcement learning (RL) approaches~\cite{wang2021deeptrader,lee2021maps,niu2022metatrader,kim2023hadaps}.
One of the most recent models, 
DeepTrader~\cite{wang2021deeptrader}, introduced an RL-based framework that consists of an asset scoring unit, market scoring unit, and portfolio generator.
Particularly, the purpose of the market scoring unit was to help the framework exploit \textit{current} market conditions in adjusting its investment decisions.
However, in real-world situations, unanticipated market dynamics may inflict severe losses in investment profits~\cite{kumbure2022forecasting_review}.
This necessitates the incorporation of market forecasts for minimizing expected risks in making asset-wise investment decisions (i.e., long and short), which naturally leads to the maximization of investment returns.
Therefore, our work establishes cross-talk between two intertwined tasks which are \textbf{market forecasting} and \textbf{optimal portfolio selection}, posing another question of how to devise an integration of each of its approaches where the former employs \textbf{reinforcement learning}.

Time-series forecasting models, mostly optimized by \textbf{supervised learning}, have undergone a series of model advancements. 
Their core design approach revolves using Transformers, which motivated many researchers to utilize them in stock price prediction tasks~\cite{malibari2021predicting,muhammad2023transformer,sridhar2021multi,wang2022stock,hu2021stock}.
Meanwhile, other Transformer variants such as FEDformer demonstrated their success in long-term time series data prediction~\cite{chen2021autoformer,zhou2021informer,zhou2022fedformer} which are attributed to their exploitation of external augmentations such as time series decomposition in Fourier domain.

Our work's central motivation lies in the integration of two related tasks associated with distinct learning strategies which are \textbf{market forecasting with supervised learning} and \textbf{portfolio selection with RL}. 
Furthermore, we present two critical research questions regarding the \textbf{\textit{under-explored compatibility of two learning methods}} and \textbf{\textit{well-known difficulty of optimizing large-scale Transformers}}.
To solve the former, we propose a transfer learning approach, by incorporating a pre-trained forecasting Transformer into a RL-based framework specializing in portfolio selection.
This approach can be facilitated by RL, characterized by formulating a reward function optimized through investment trials under realistic and different market conditions.

However, the latter remains a challenge as most state-of-the-art forecasting Transformers have massively large amount of trainable parameters especially the FEDformer (approximately 60 million in our case)~\cite{corberan2023portfolio_forecasting_genetic}.
While fine-tuning is deemed the primary choice in optimizing models for specific downstream tasks, adjusting a large number of parameters incurs additional computational burden.
To cope with this, Low-Rank Adaptation (LoRA) were proposed to efficiently apply transfer learning from the pre-trained weights to downstream tasks~\cite{hu2021lora}.
Moreover, LoRA exhibited performance enhancements comparable to full fine-tuning methods, showing success in vision-to-text~\cite{ye2023lora_vision_text} and text-to-text~\cite{wang2024prolificdreamer_lora_3d, alpaca-lora} transfer learning works.

Perhaps the most motivational relevant work in our study design \cite{ouyang2022rlhf}'s approach to adopt RL in training large language models, mainly Transformer-based, on downstream tasks related to natural language processing.
In fact, this work highlights LoRA's transferibility of textual knowledge trained with supervised learning method to task adaptation based on RL~\cite{sun2023aligning_text_to_rl}.
Therefore, we hypothesize that utilizing this approach in integration of two methodologies used in \textbf{market forecasting} and \textbf{optimal portfolio selection} task can lead to improved investment outcomes, which aligns with the common notion of market outlooks being useful in devising optimal portfolio selection strategies.

In this work, we introduce \textbf{\textit{\textsc{DeepClair}, a deep clairvoyant portfolio selection model that leverages not only its current understanding of the market situation but also its forecasting of future market states}}.
\textsc{DeepClair} leverages the advanced deep RL framework capabilities and incorporates a recently developed Transformer-based model that views time-series market data in the frequency domain.
To the best of our knowledge, our novelty lies in the attempt to incorporate current transformer-based market forecasting into RL-based portfolio selection.
This work represents a cornerstone empirical trial demonstrating that supervised pre-training tasks can aid RL scenarios, which are challenging to optimize in real-world environments.
The contributions in this work are the following:

\begin{itemize}
    \item To the best of our knowledge, this is the first attempt to incorporate the current transformer-based market forecasting methodology into an automatic portfolio selection framework.
    \item We pre-trained recently proposed transformer-based models on various time-series datasets and empirically chose FEDformer with LoRA adaptation as our market price forecasting module.
    \item Experiments on portfolio selection demonstrate \textsc{DeepClair}'s ability to capitalize on its market forecasting ability to achieve the best investment returns compared to its baselines.
    Moreover, qualitative analysis shows how \textsc{DeepClair} can adaptively shift its positioning strategy during two phenomenal bullish and bearish periods.
\end{itemize}

\section{Related Works}

\subsection{Time-series Forecasting Transformers}
The evolution of time-series forecasting models witnessed a significant transition from earlier approaches to the advent of transformers. Informer~\cite{zhou2021informer} represents a notable model in this shift, followed by more recent advancements like Autoformer~\cite{wu2021autoformer} and FEDformer~\cite{zhou2022fedformer}. 
These newer models capitalize on advanced capabilities for composing seasonal trends.

Traditional portfolio strategies often lack adaptability to dynamic market conditions. 
In response, the introduction of RL-based portfolio selection models, as discussed in earlier works like DeepTrader~\cite{wang2021deeptrader} and HADAPS~\cite{kim2023hadaps}, has aimed to address this limitation.

Transformer-based time-series forecasting systems demonstrated success in various domains~\cite{zhang2023flight_transformer_timeseries_example1,jiang2023traffic_transformer_timeseries_example2,peng2022tlt_water_quality_transformer_timeseries_example3}.
However, within our current understanding, attempts to integrate forecasting into portfolio selection approaches have primarily utilized deterministic models, as seen in the use of genetic algorithms~\cite{corberan2023portfolio_forecasting_genetic}.

\subsection{Reinforcement Learning adopting to Downstream Tasks}
Adapting to downstream tasks utilizing the RL method was originally discussed in the context of traditional RL problems such as robotics~\cite{christiano2017rlhf_original1}. 
It has since demonstrated its utility in the textual domain by adapting pre-trained language models to downstream tasks such as text summarization~\cite{stiennon2020rlhf_original2}. 
Furthermore, research by ~\cite{ouyang2022rlhf} has shown that language model training can adapt effectively to smaller human relevance corpora. 
Notably, the success of models like LLaMA2\cite{touvron2023llama2} has further validated the efficacy of this mechanism.

\subsection{Parameter Efficient Fine-Tuning Method}

After the Adapter~\cite{houlsby2019adapter} method came out to transfer the knowledge of the pre-trained textual language model to downstream text classification tasks, several methods have been introduced, including soft prompts~\cite{lester2021softprompt}, IA3~\cite{liu2022ia3}, and OFT~\cite{liu2023oft}. 
Among these, LoRA~\cite{hu2021lora} has emerged as a prominent approach.

The LoRA method has shown its possibility not only in textual adaptation tasks such as chatbot~\cite{alpaca-lora} and recommendation systems with textual cues~\cite{bao2023lora_text_tallrec} but also in vision-related domains.
These include Vision Question Answering (VQA)~\cite{keita2024bilora_lora_vqa, ye2023lora_vision_text}, 2D to 3D image generation~\cite{liu2024lora_vision_2d_to_3d} and text to 3D generation~\cite{wang2024prolificdreamer_lora_3d}.
Additionally, LoRA has been explored for textual alignment using RLHF~\cite{ouyang2022rlhf}, showing potential for adapting models trained with supervised learning to RL context~\cite{sun2023aligning_text_to_rl}.

\begin{figure*}
    \centering
    \includegraphics[width=0.7\textwidth]{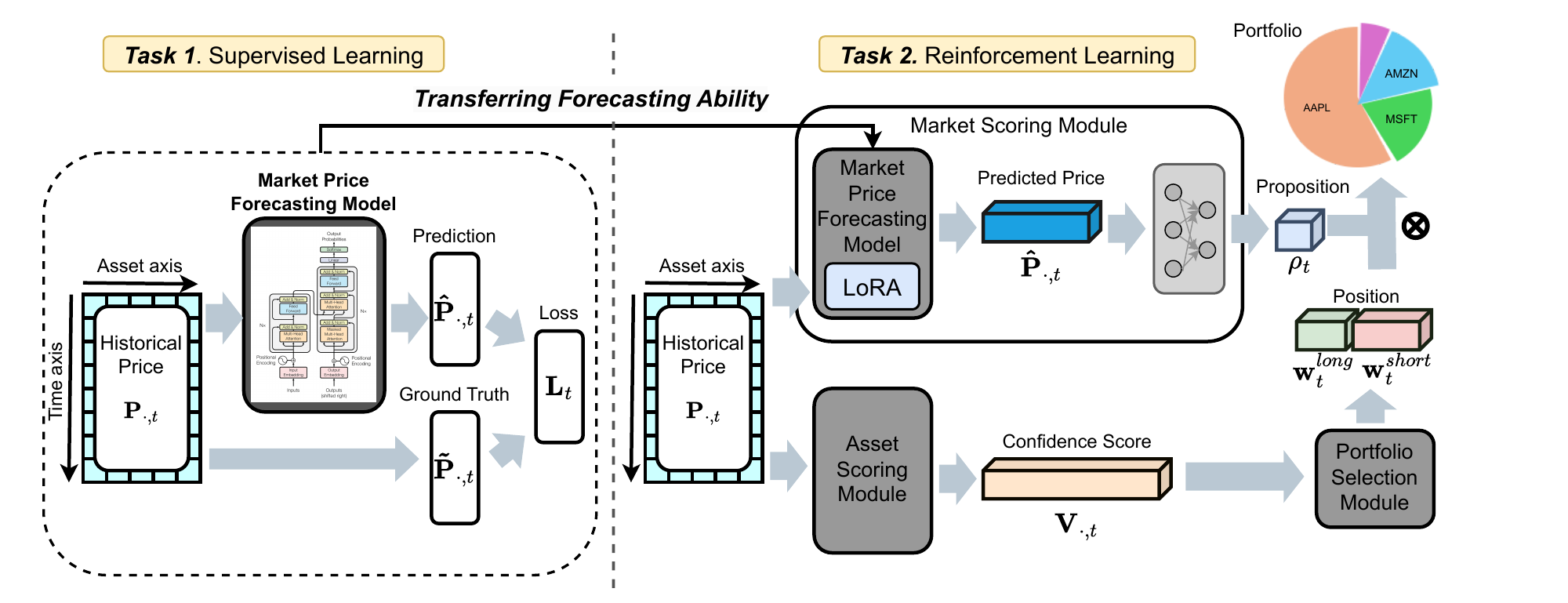}
	\caption{
The model architecture of \textsc{DeepClair} is detailed in Sec~\ref{sec:deepclair}. 
Task 1 involves pre-training the forecasting module using a supervised learning approach using Market Price Forecasting Model $M_1$ described in Sec~\ref{subsec:market_price_forecasting_model}. 
In Task 2, we use the forecasting model \(M_1\) with pre-trained weights from Task 1, fine-tuning a relatively small number of parameters with LoRA techniques. 
During investment trials through RL, this fine-tuned model is used for the portfolio selection task, defined as a portfolio optimization framework \(M_2\) described in Sec~\ref{subsec:portfolio_optimization_framework}.
The portfolio selection module generates long $\mathbf{w}_t^{long}$ and short $\mathbf{w}_t^{short}$ position vectors with the outputs $\mathbf{v}_t$.
Based on the forecasting outcomes, \textsc{DeepClair} determines the long and short ratios of each asset using the vector $\rho_t$. 
The actual portfolio ratios for each asset by combining the outputs $\mathbf{w}_t$ from the Portfolio Selection Module and $\rho_t$ from the Market Scoring Module.
	}
	\label{fig:method_fig}
\end{figure*}

\section{Problem Statement}


The following details are the preliminaries for our tasks.

\begin{problem}[Optimal Portfolio Selection]
\label{problem:portfolio-selection}
Given a set of $n$ assets $\mathcal{A}=\{a_1, a_2, \ldots, a_n\}$, we denote each $i$th asset $a_i$ having historical stock prices as $\mathbf{P}_{i,t}=[\mathbf{p}_{i,t-w},\cdots,\mathbf{p}_{i,t-1}]$ where $\mathbf{p}_{i,j}$ is the closing market price of asset $a_i$ at historical time step $j$. 
At current time step $t$ prior to market closing and given a window size of historical prices $w$ ($w\geq1$), the objective of \textsc{DeepClair} is to dynamically adjust the portfolio composition $\mathcal{C}=[\mathbf{c}_{1,t},\mathbf{c}_{2,t},\cdots,\mathbf{c}_{n,t}]$ where $\mathbf{c}_{i,t}$ is the portfolio ratio of asset $a_i$ at time $t$.
The goal is to maximize the return $\delta_t = \sum^{n}_{i=1}\delta_{i,t}c_{i,t}$, where $\delta_{i,t}=(\mathbf{p}_{i,t}-\mathbf{p}_{i,t-1})/\mathbf{p}_{i,t-1}$.
\end{problem}

In sum, the goal of a system is to make adaptive investment decisions given dynamic market circumstances at time $t$.
As illustrated above, market forecasts can contribute to refining portfolio strategies for better investment returns.
To solve \ref{problem:portfolio-selection}, we formulate two tasks which are \textbf{Learning to Forecast Market} and \textbf{Learning Optimal Portfolio Selection using the Transferred Market Forecasting Module}.

\begin{task}[Learning to Forecast Market] 
\label{task:pre-training}
For each $i$th asset $a_i\in\mathcal{A}$ and discrete time step $t$, the agent's task is to predict its future market state $\mathbf{\hat{P}}_{i,t}=[\mathbf{\hat{p}}_{i,t},\cdots,\mathbf{\hat{p}}_{i,t+h-1}]$ of forecast horizon $h$ ($h\geq1$) given its historical prices $\mathbf{P}_{i,t}=[\mathbf{p}_{i,t-w},\cdots,\mathbf{p}_{i,t-1}]$ of window size $w$. The learning objective in this task is to minimize forecast error between $\mathbf{\hat{P}}_{i,t}$ and $\mathbf{\tilde{P}}_{i,t}$ where $\mathbf{\tilde{P}}_{i,t}=[\mathbf{\tilde{p}}_{i,t},\cdots,\mathbf{\tilde{p}}_{i,t+h-1}]$ is the actual sequence of $h$ closing market prices for asset $a_i$. 
\end{task}
Task~\ref{task:pre-training} serves as an auxiliary task, aiding the system in understanding and adapting to market trends, thereby improving its ability to make informed investment decisions.
 

\begin{task}[Learning Optimal Portfolio Selection using the \textbf{Transferred} Market Forecasting Model] 
\label{task:fine-tuning}
The agent's task is to formulate an optimal portfolio strategy \(\mathcal{C}\in\mathbb{U}\) for all assets \(a_i\in\mathcal{A}\) at discrete time step \(t\). 
This involves considering both the actual market data \(\mathbf{P}_{i,t}\) and the predicted market data \(\mathbf{\hat{P}}_{i,t}\) to maximize the total return \(\delta_t\). 
\(\mathcal{C}_t\) represents the vector of allocation weights \(c_{i,t}\) for each asset \(a_i\), indicating the proportion of the total portfolio value invested in each asset at time \(t\).
\end{task}


Task~\ref{task:fine-tuning} and Task~\ref{task:pre-training} are \textsc{DeepClair}'s primary and auxiliary tasks respectively. Task~\ref{task:pre-training} contributes to improving \textsc{DeepClair}'s understanding in market trends and ability to exploit them in investment decisions involved in Task~\ref{task:fine-tuning}.

\section{\textsc{DeepClair} }
\label{sec:deepclair}

\textsc{DeepClair} is an advanced portfolio selection system that combines two components described in Fig~\ref{fig:method_fig}: 1) a transformer-based time-series forecasting model $M_1$ for portfolio evaluation to solve Task~\ref{task:pre-training} and 2) a RL-based portfolio optimization framework $M_2$ to solve Task~\ref{task:fine-tuning}.

\subsection{Market Price Forecasting Model}
\label{subsec:market_price_forecasting_model}

The primary objective of this framework is to predict future asset prices $\mathbf{\hat{P}}_{\cdot,t}$ based on a given market signal $\mathbf{P}_{\cdot,t}$. 

\begin{align}
	\label{eq:forecasting_unit}
    \mathbf{\hat{P}}_{\cdot,t} &=  M_{1} (\mathbf{{P}}_{\cdot,t}) 
\end{align}
where $\mathbf{\hat{P}}_{\cdot,t} \in \mathcal{R}^{n} $ and $\mathbf{{P}}_{\cdot,t} \in \mathcal{R}^{n,w}$, $w$ is a window size of historical prices and $n$ represents the number of assets. 

In this study, we adopted FEDformer~\cite{zhou2022fedformer} as our forecasting model $M_1$, employing 2 encoder blocks and 1 decoder block. 
FEDformer is renowned for its ability to capture the global view of time-series data, including overall trends, through a seasonal-trend decomposition method. 
This aligns with our model's objective of predicting preceding prices by capturing broader market trends, aiding in determining the proportion of long and short positions.

\subsubsection{Training Procedures by Time Series Forecasting}

We optimize $M_1$ as regression problem for solving Task~\ref{task:pre-training} by minimizing the prediction of the Forecasting model $\mathbf{\hat{P}}$ and the real price data $\mathbf{\tilde{P}}$.

\subsection{Portfolio Optimization Framework}
\label{subsec:portfolio_optimization_framework}
The objective of this framework is to optimize a portfolio optimization framework $M_2$ by a RL method.
$M_2$ consists of Asset Scoring Module, Portfolio Selection Module, and Market Scoring Module. 

\subsubsection{Asset Scoring Module}
The asset scoring module produces a confidence vector $\mathbf{v}_{\cdot,t} = \{v_{1,t},\cdots,v_{a,t},\cdots,v_{n,t} \}$, where $v_{a,t}$ is a confidence score of asset $a$ at time $t$ and $n$ represents the number of assets.

\begin{align}
	\label{eq:asset_scoring_unit}
    \mathbf{v}_{\cdot,t} &=  ASM(\mathbf{{P}}_{\cdot,t})
\end{align}

The notation of ASM denotes the Asset Scoring Module, with a window size of historical input prices $w$ set to 5.
The asset scoring unit integrates temporal and graph convolution layers, along with a spatial attention mechanism, enabling the capture of temporal correlations among asset prices.

\subsubsection{Portfolio Selection Module}
The portfolio selection module's (PSM) role is selection of long and short position vectors, $\mathbf{w}^{long}_t$ and $\mathbf{w}^{short}_t$, respectively.
Here, $\mathbf{w}_t = \{w_{1,t},w_{2,t}, \cdots, w_{n,t}\}$, where $w_{a,t}$ is a weight for long or short position for asset $a$.

The PSM selects top $n_{long}$ and a list of its indexes as $\mathbf{v}^{long}_t$ and for bottom $n_{short}$ assets, $\mathbf{v}^{short}_t$ for each portfolio selection by using the asset score vector $\mathbf{v}_t$.
Then PSM decides the proportions of each top and bottom assets with the proportion vector $\mathbf{w}^{long}_t$
and $\mathbf{w}^{short}_t$.

\begin{align}
    w_{i,t}^{long} = 
    \begin{cases}
        \frac{exp(v_{i,t})}{\sum_{j\in \mathbf{v}_t^{long}} exp(v_{j,t})} & \text{if } i \in \mathbf{v}^{long}_t \\
        0 & \text{if } i \notin \mathbf{v}^{long}_t 
    \end{cases} 
    \label{eq:long_weight}
\end{align}

\begin{align}
    w_{i,t}^{short} = 
        \begin{cases}
        \frac{- exp(1 - v_{i,t})}{\sum_{j\in \mathbf{v}^{short}_t} exp(1 - v_{j,t})} & \text{if } i \in \mathbf{v}_t^{short} \\
        0 & \text{if } i \notin \mathbf{v}_t^{short}
    \end{cases}
    \label{eq:short_weight}
\end{align}
each elements in long $\mathbf{w}^{long}_t$ and short $\mathbf{w}^{short}_t$ vectors are normalized with the equations~\ref{eq:long_weight} and \ref{eq:short_weight}.
We apply the softmax function for selecting portfolio actions because using the softmax function to optimize discrete action spaces in RL is a well-established approach~\cite{sutton2018rlbook, mnih2016softmax1, schulman2015softmax2}.
This method is particularly relevant in the context of portfolio selection with RL~\cite{wang2021deeptrader, kim2023hadaps, lee2021maps}.
A value of \(0\) is assigned to the assets for which no position is selected.

\subsubsection{Market Scoring Module}

Since market conditions are inherently unpredictable and uncertain~\cite{deng2016uncertain_market_deep}, it is impossible to accurately predict stock movements solely based on historical data.
However, previous approaches attempted to utilize the current market conditions to select actions aimed at achieving future optimal positions~\cite{wang2021deeptrader}, under the naive assumption that the current market status will persist~\cite{markowitz1959portfolio}. 
We advance this approach by incorporating outputs from the forecasting module to better inform decision-making.

The Market Scoring Module (MSM) is responsible for determining $\rho_t$, which represents the Long and Short proposition of \textsc{DeepClair}. 

\begin{align}
   \mathbf{\hat{P}}_{\cdot,t} &= M_1(\mathbf{{P}}_{\cdot,t}) \\
MSM(\mathbf{{P}}_{\cdot,t}) &= \text{Linear}(M_1(\mathbf{{P}}_{\cdot,t})) \\
  \mu_t, \sigma_t &= MSM(\mathbf{{P}}_{\cdot,t})
\end{align}

During the training phase, we sample $\rho_t$ from the distribution $N(\mu_t, \sigma_t)$. 
In testing, we use $\mu_t$ as the value of $\rho_t$.
This technique is for stable exploring the agent's action space, widely used in RL domains after Soft Actor-Critic technique~\cite{haarnoja2018sac}, recently in the portfolio selection domain~\cite{wang2021deeptrader,kim2023hadaps}.

\paragraph{Adaptation of Forecasting Module using LoRA}
We use LoRA\cite{hu2021lora} to adapt the forecasting module $M_1$, which was trained in the pre-training task (Task~\ref{task:pre-training}), to the fine-tuning task (Task~\ref{task:fine-tuning}). The primary objectives of integrating the LoRA module are twofold: 1) to retain the forecasting ability established during the initial pre-training phase; and 2) to adapt the module's behavior for portfolio selection during the fine-tuning stage.
We apply LoRA adapters to every transformer block in the encoders of forecasting model $M_1$. 

\subsubsection{Formulating Portfolio Selection Vectors}
\label{portfolio_selection_vector}
Final portfolio selection is done by using the proportion $\rho_t$ between the long position vector $\mathbf{w}_{i,t}^{long}$ and we note that the short position vector $\mathbf{w}_{i,t}^{short}$.

\begin{align}
& \mathbf{Portfolio}^{long}_t = \rho_t \times \{w^{long}_{0,t}, w^{long}_{1,t}, \cdots w^{long}_{n,t} \} \\
& \mathbf{Portfolio}^{short}_t = (1-\rho_t) \times \{w^{short}_{0,t}, w^{short}_{1,t}\cdots w^{short}_{n,t} \} 
\end{align}


\subsection{Training Procedures by Portfolio Selection}

The primary objective of \textsc{DeepClair} is to maximize reward $r_1$ by exploring the action spaces defined in investment outcome $\mathbf{Portfolio}^{long}$ and $\mathbf{Portfolio}^{short}$.
This $r_1$ is defined as :

\begin{align}
r_{1,t} &= \sum ( \delta_t \circ \mathbf{Portfolio}^{long}_t + \delta_t \circ \mathbf{Portfolio}^{short}_t )
\end{align}
where $\delta_t$ is a rate of return that is calculated by price change within the target day and $\circ$ stands for element-wise multiplication.
If the directions of the price change and the position are the same, $r_{1,t}$ has a positive value that strengthens the behavior by the value of amplifications, and if not, $r_{1,t}$ acts like a negative reward because it has a negative value.
The output is connected to the decision of the Asset Scoring Module and the Forecasting Module, so $r_{1,t}$ is a signal to update all the parameters in Asset Scoring Module and the LoRA parameters in the Forecasting Module.

Also, we add an auxiliary reward $r_{2,t}$ to maximize the prediction capability of the Asset Scoring Module by using all of the confidence prediction parameterized as $\mathbf{v}_t$.
\begin{align}
r_{2,t} &= \sum \delta_t \circ \mathbf{v}_{\cdot, t} 
\end{align}
$r_1$ uses the final output of \textsc{DeepClair}, which is straightforward, but this neglects the other confidence scores in $\mathbf{v}$, which could be helpful signals while training.
$r_2$ is defined to overcome the shortage of formulation about $r_1$.

The final training objective of \textsc{DeepClair} is defined as follows:
\begin{align}
J(\phi) &= \mathbb{E}_{ \mathbf{v_{\cdot,t}} } [ \alpha   log(\rho_t) \times r_{1,t} + \beta r_{2,t}  ] 
\end{align}
 $\beta$ are used to control the magnitude of the $r_{2,t}$.
 We set $\alpha$ as 0.05, $\beta$ as 1. 

\section{Experimental Settings}

\subsection{Dataset and Evaluation}

We evaluated our model using Nasdaq and Dow Jones datasets, selecting the top 26 stocks by market capitalization as of 2023, with data from January 1992 to the present, sourced from Yahoo Finance\footnote{\url{https://finance.yahoo.com/}}. 
Details are in Table~\ref{table:data_split}. For Task~\ref{task:pre-training}, the development and test datasets were split over the same period. In Task~\ref{task:fine-tuning}, the test period was extended for a more robust evaluation.

Additionally, we used two types of datasets to evaluate the performance of our forecasting model (Task~\ref{task:pre-training}). 
The first dataset, from existing literature~\cite{zhou2021informer}, includes the ETTh1 (Electricity Transformer Temperature) dataset with 1-hour-level load and oil temperature data spanning 2 years, and an electricity dataset with hourly consumption data of 321 clients over 2 years. 
    The second dataset category includes our Nasdaq and Dow Jones datasets, used to evaluate the model's pricing data predictions.

All experiments were repeated ten times with different random seeds and their results are based on mean performance.
The evaluation measures to evaluate investment performance that were used in the experiments are the following:
Annualized Return (ARR), Annualized Sharpe Ratio (ASR), Maximum Drawdown (MDD), Calmar Ratio (CR), and Sortino Ratio (SoR).
ARR measures the overall return of a method, ASR, SoR, and CR evaluate risk-adjusted returns, and MDD assesses risk.

\begin{table}[]
\scalebox{0.7}{
\begin{tabular}{llll}
\toprule
Task &  Train      & Dev       & Test                 \\
\midrule
\ref{task:pre-training}  &  1992.01$\sim$2004.09 & 2004.10 $\sim$ 2006.05 &  2006.06 $\sim$ \textbf{2008.12} \\
\ref{task:fine-tuning}  & 1992.01$\sim$2004.09 & 2004.10 $\sim$ 2006.05 & 2006.06 $\sim$ \textbf{2022.12}\\
\bottomrule
\end{tabular}
} 
\caption{
Details of the experimental dataset.
}
\label{table:data_split}
\end{table}


\begin{table*}[]
\centering
\scalebox{0.65}{
\begin{tabular}{l rrrrrr rrrrrr}

\toprule
 & \multicolumn{6}{c}{\textbf{Nasdaq}} & \multicolumn{6}{|c}{\textbf{Dow Jones}} \\
 & \multicolumn{1}{c}{\textbf{ARR}(\%)$\uparrow$} & \multicolumn{1}{c}{\textbf{ASR$\uparrow$}} & \multicolumn{1}{c}{\textbf{AVol$\downarrow$}} & \multicolumn{1}{c}{\textbf{MDD}(\%)$\downarrow$} & \multicolumn{1}{c}{\textbf{CR$\uparrow$}} & \multicolumn{1}{c}{\textbf{SoR$\uparrow$}} & \multicolumn{1}{|c}{\textbf{ARR}(\%)$\uparrow$} & \multicolumn{1}{c}{\textbf{ASR$\uparrow$}} & \multicolumn{1}{c}{\textbf{AVol$\downarrow$}} & \multicolumn{1}{c}{\textbf{MDD}(\%)$\downarrow$} & \multicolumn{1}{c}{\textbf{CR$\uparrow$}} & \multicolumn{1}{c}{\textbf{SoR$\uparrow$}} \\

\midrule

Benchmark Index
&9.77&0.434&0.225&55.63 &0.176 &6.436& 6.71 & 0.348 & 0.192 & 53.78 & 0.125 & 5.239 \\
\midrule
CSM& 2.54&0.158&0.161& 48.18 &0.053& 2.294&  1.02 & 0.077 & \textbf{0.132} & 39.98 & 0.025 & 1.091 \\ 
BLSW& -4.95&-0.308&0.161& 71.07 &-0.070& -5.325& -2.72 &-0.207&\textbf{0.132}& 57.02 &-0.048&-3.713\\
\midrule
EIIE (2017) & 6.51 & 0.369& 0.176 & 41.15 & 0.185   &5.930&4.54&0.310	& 0.147	& 41.08&0.119&4.898  \\
HADAPS (2023) & 8.92 & 0.575 & \textbf{0.157} & 38.46 & 0.244 & 8.851&5.64&0.392	& 0.142 &\textbf{39.28}	&0.153	&6.039\\
DeepTrader (2021) &11.81 & 0.619 & 0.191 & \textbf{37.04} & 0.326& 9.575 &8.47	&0.479	&0.176	& 44.94	&0.188&7.420\\
\midrule
\textbf{\textit{DeepClair}} & \textbf{15.22} & \textbf{0.659} & 0.231 & 48.59 & \textbf{0.346} & \textbf{10.593} &\textbf{10.30}	&\textbf{0.526}	& 0.196	& 46.77	&\textbf{0.220}	&\textbf{8.052} \\
\bottomrule

\end{tabular}
}
\caption{
The results in different stock markets.
}
\label{table:main}
\end{table*}


\subsection{Model Baselines}
The baseline models used in the experiments and categorized by their approaches are the following,
\begin{enumerate}
    
    \item \textbf{Benchmark Index}: This indicates the market trends of each of its associated major index in the U.S. market.
   
    \item \textbf{Traditional Investment Strategies}: Cross-Sectional Momentum strategy(CSM)~\cite{jegadeesh1993returns} and Buy Losers Sell Winners strategy(BLSW)~\cite{debondt1987further}.
    

\item \textbf{Deep Reinforcement Learning Approaches}\footnote{We select RL-based portfolio selection  systems for day-trading (exclude intra-day trading~\cite{sun2022deepscalper}) and open-sourced implementations.}:
\textit{EIIE} (\textit{Ensemble of Identical Independent Evaluators})~\cite{jiang2017eiie} is a foundational open-source RL framework for portfolio selection. \textit{DeepTrader}~\cite{wang2021deeptrader} and \textit{HADAPS} (\textit{Hierarchical Adaptive Multi-Asset Portfolio Selection})~\cite{kim2023hadaps} are recent RL-based portfolio management systems.

\end{enumerate}

\begin{table*}[]
\centering
\scalebox{0.65}{
\begin{tabular}{l rrr rrr rrr rrr}
\toprule
 & \multicolumn{3}{c}{\textbf{ETTh1}} & \multicolumn{3}{|c}{\textbf{Electricity}} &\multicolumn{3}{|c}{\textbf{NASDAQ}}&\multicolumn{3}{|c}{\textbf{Dow Jones}}\textbf{}\\
 &  \multicolumn{1}{c}{\textbf{MAE}$\downarrow$} & \multicolumn{1}{c}{\textbf{RMSE}$\downarrow$} & \multicolumn{1}{c}{\textbf{MAPE}$\downarrow$} &  \multicolumn{1}{|c}{\textbf{MAE}$\downarrow$} & \multicolumn{1}{c}{\textbf{RMSE}$\downarrow$} & \multicolumn{1}{c}{\textbf{MAPE}$\downarrow$}  &  \multicolumn{1}{|c}{\textbf{MAE}$\downarrow$} & \multicolumn{1}{c}{\textbf{RMSE}$\downarrow$} & \multicolumn{1}{c}{\textbf{MAPE}$\downarrow$} &  \multicolumn{1}{|c}{\textbf{MAE}$\downarrow$} & \multicolumn{1}{c}{\textbf{RMSE}$\downarrow$} & \multicolumn{1}{c}{\textbf{MAPE}$\downarrow$}  \\

\midrule

Transformer      &0.387 &0.564 &9.677 &0.320 &0.448 &3.018&2.484 &3.912 &0.826&1.577 &3.098 &0.543 \\
Informer         &0.437 &0.634 &9.340 &0.377 &0.516 &3.744&2.579 &4.040 &0.873&1.670&3.169 &0.852    \\
Autoformer       &0.394 &0.573 &10.158 &0.275 & \textbf{0.386} &2.449&0.452 &0.775 &\textbf{0.041}&0.286 &0.476 & \textbf{0.037}    \\
\textbf{FEDformer} &\textbf{0.339} &\textbf{0.493} &\textbf{8.841} & \textbf{0.273} & \textbf{0.386} & 
\textbf{2.428 }& \textbf{0.445} & \textbf{0.759} & \textbf{0.041}&\textbf{0.281} &\textbf{0.467} &\textbf{0.037}      \\ 
\bottomrule

\end{tabular}
}
\caption{Comparison of forecasting modules. Baseline forecasting modules are vanilla Transformer, Informer, and Autoformer.}
\label{table:sub}
\end{table*}

\begin{table*}[]
\centering
\scalebox{0.65}{
\begin{tabular}{lrrrrrr rrrrrr}
\toprule
 & \multicolumn{6}{c}{\textbf{Nasdaq}} & \multicolumn{6}{|c}{\textbf{Dow Jones}} \\
 & \multicolumn{1}{c}{\textbf{ARR}(\%)$\uparrow$} & \multicolumn{1}{c}{\textbf{ASR}$\uparrow$} & \multicolumn{1}{c}{\textbf{AVol$\downarrow$}} & \multicolumn{1}{c}{\textbf{MDD}(\%)$\downarrow$} & \multicolumn{1}{c}{\textbf{CR$\uparrow$}} & \multicolumn{1}{c}{\textbf{SoR$\uparrow$}} & \multicolumn{1}{|c}{\textbf{ARR}(\%)$\uparrow$} & \multicolumn{1}{c}{\textbf{ASR$\uparrow$}} & \multicolumn{1}{c}{\textbf{AVol$\downarrow$}} & \multicolumn{1}{c}{\textbf{MDD}(\%)$\downarrow$} & \multicolumn{1}{c}{\textbf{CR$\uparrow$}} & \multicolumn{1}{c}{\textbf{SoR$\uparrow$}} \\

\midrule
\multicolumn{1}{c}{\textbf{\textit{Ablations on Forecasting Module}}}  \\ 
Transformer-LoRA-Encoder & 8.10 & 0.358& 0.214 & 45.80 & 0.193&5.634   &5.03&0.265	&0.178	& 43.77 &0.117&4.261\\
Informer-LoRA-Encoder & 6.20 & 0.282& 0.206 & 46.16 & 0.150   &4.444&3.56&0.194	&0.168	& 45.03&0.087&3.108 \\
Autoformer-LoRA-Encoder & 8.76 & 0.354& 0.227 & 47.00 & 0.204   &5.523&5.74&0.290	&0.180	& 44.98&0.130&4.610 \\
		
\midrule
\midrule
\multicolumn{1}{c}{\textbf{\textit{Ablations on optimization of FEDformer}}}  \\ 
FED-Removed & 6.97 & 0.354 & \textbf{0.198} & 48.37 & 0.152 & 5.620 &4.67	&0.282	& \textbf{0.167}	& \textbf{40.96}	&0.120	&4.612 \\
FED-Frozen& 11.01 & 0.436& 0.240 & 46.73 & 0.249   &6.813&5.39&0.254	&0.188	& 48.53&0.120&4.031 \\
FED-Finetuning & 7.61 & 0.313& 0.224 & 45.88 & 0.173   &4.902&5.69&0.275	&0.188	& 44.64&0.127&4.352 \\
\midrule

\multicolumn{1}{c}{\textbf{\textit{Ablations on LoRA applied to FEDformer}}}  \\ 
FED-LoRA-Decoder &12.58 & 0.577 & 0.246 & 50.42 & 0.256 & 7.851 &5.25	&0.256	& 0.182&46.14&0.120&4.069\\
FED-LoRA-FEA  & 8.49 & 0.359& 0.224 & 46.89 & 0.187  &5.629 &4.36&0.173	&0.190	& 58.39&0.119&2.746 \\
FED-LoRA-All  &  11.16& 0.444 & 0.238 & \textbf{45.62} & 0.254   &6.932&4.84&0.240	&0.182	& 45.85&0.100&3.778 \\

\midrule
\textbf{\textsc{DeepClair}} (FED-LoRA-Encoders)  & \textbf{15.22} & \textbf{0.659} & 0.231 & 48.59 & \textbf{0.346} & \textbf{10.593} &\textbf{10.30}	&\textbf{0.526}	& 0.196	& 46.77	&\textbf{0.220}	&\textbf{8.052} \\
\bottomrule
\end{tabular}
}
\caption{
Ablation studies on employing LoRA to \textsc{DeepClair}. 
Note that \textsc{DeepClair} applies LoRA exclusively to encoders, setting the rank value at 4,000 (representing about 75\% of the pre-trained FEDformer encoders in \textsc{DeepClair}). 
The rank value of 4,000 remains consistent across ablations involving LoRA on other FEDformer components (Decoder, FEA, and All).
}
\label{table:forecasting}
\end{table*}

\section{Experimental Results}
\textsc{DeepClair} was designed as a RL-based portfolio selection AI system, harnessing the predictive capabilities of a current transformer-based time-series forecasting model.
We conducted experiments to address the following research questions, validating the underlying design principles:

\textbf{RQ1.} Does incorporation of market forecasting into portfolio selection framework contribute to better investment results?

\textbf{RQ2.} For architectural design of \textsc{DeepClair}, what is the optimal market forecasting transformer-based model? 

\textbf{RQ3.} For incorporation of FEDformer into \textsc{DeepClair}, what is the best optimization strategy?

\textbf{RQ4.} Can we find empirical evidence supporting \textsc{DeepClair}'s adaptive capabilities in capturing market shifts?




\subsection{RQ1. \textsc{DeepClair} shows superior investment performance and greatly benefits from market forecasts}
Table~\ref{table:main} shows the results of the main experiments conducted on two separate datasets.
The purpose of these experiments was to comprehensively evaluate and compare \textsc{DeepClair}'s portfolio selection with other baselines. 
Notably, \textsc{DeepClair} outperformed its baselines in most evaluation metrics including \textbf{ARR}, \textbf{ASR}, \textbf{CR} and \textbf{SoR} in both datasets Nasdaq (15.22, 0.659, 0.346, 10.593) and Dow Jones (10.30, 0.526, 0.220, 8.052).
The performance gain in terms of investment return is significant based on a comparison with the other baseline methods on both datasets.

According to the experimental results, traditional investment strategies such as \textit{CSM} and \textit{BLSW} underperformed across all evaluation metrics on both datasets. 
This indicates that during our testing period, traditional strategies that follow trends (CSM) or their reverse (BLSW) were less effective compared to more adaptive trading strategies.
\textsc{DeepClair} preemptively capitalizes on its forecasting ability to make rational investment decisions and evade the severe drawbacks of swaying itself towards uncertain market trends.

The experimental results featuring deep RL model baselines imply that adaptive investment strategies help mitigate risks.
This is empirically shown by \textsc{DeepClair} unable to show the best performance in terms of risk measures \textbf{AVol} and \textbf{MDD} in both datasets (0.231, 48.59 and 0.196, 46.77).
However, \textsc{DeepClair} which also employs deep RL to facilitate adaptive portfolio selection, still outperforms in other evaluation metrics related to risk-adjusted returns (\textbf{CR} and \textbf{SoR}).
This suggests that integrating adaptive investment strategies with studying market trends amplifies portfolio selection's effectiveness.


	

\subsection{RQ2. FEDformer is a suitable choice for the Market Scoring Module in \textsc{DeepClair}}
\label{subsection:rq2-fedformer-is-suitable}
Prior to choosing which transformer-based model is suitable for our \textsc{DeepClair} framework, we conducted experiments on four different time series datasets which are ETTH1, Electricity, Nasdaq and Dow Jones in Table~\ref{table:sub}. 
We first selected previously introduced forecast models which are \textit{Transformer}, \textit{Informer}, \textit{Autoformer}, and \textit{FEDformer}.
A clear distinction in terms of forecasting performance (\textbf{MAE}, \textbf{RMSE}, and \textbf{MAPE}) was found between \textit{Transformer}, \textit{Informer} and \textit{Autoformer}, \textit{FEDformer}. 
This difference is mostly attributed to the seasonal-trend decomposition approach exploited by both \textit{Autoformer} and \textit{FEDformer}~\cite{wu2021autoformer,zhou2022fedformer} which is an autoregressive property helpful in long time-series forecasting. 
As shown in Table~\ref{table:sub}, \textit{FEDformer} has outperformed other baselines in all three evaluation metrics. 

Additionally, we conducted an ablation study on different transformer variants used as \textsc{DeepClair}'s market price forecasting model which are Transformer, Informer, and Autoformer, as shown in Table~\ref{table:forecasting}.
In both the Nasdaq and Dow Jones datasets, FEDformer outperformed other transformer variants in providing accurate market forecasts which led \textsc{DeepClair} to make effective long and short decisions in unseen market circumstances.
This is attributed to its ability to extract meaningful seasonal patterns in the frequency domain during the pre-training phase and exploit them to foresee fourteen years of uncertain market situations in an adaptive manner.

\subsection{RQ3. Low Rank Adaptation applied to FEDformer's Encoder benefits optimization of \textsc{DeepClair}}
As shown in Table~\ref{table:forecasting}, we explored various ways of incorporating the pre-trained FEDformer model into the adopted DeepTrader-based framework. 
The FEDformer was designed based on encoder-decoder architecture where two of the modules are interconnected via the Frequency Enhanced Attention block (FEA)~\cite{zhou2022fedformer}. 
Importing pre-trained transformer-based models and fine-tuning them on different downstream tasks has been a conventional practice in other domains, especially natural language processing~\cite{hu2021lora}. 
As this is our first attempt to utilize the FEDformer in \textsc{DeepClair}, we conducted additional experiments related to several optimization methods associated with large pre-trained models. 
The additional experiments related to FEDformer are the followings\footnote{Note that \textsc{DeepClair} is equivalent to FED-LoRA-Encoder}:

\begin{itemize}
\item \textit{FED-Removed/Frozen/Finetuning}: Ablated versions of \textit{DeepClair} with FEDformer removed, parameters frozen, or parameters trainable, respectively.
\item \textit{FED-LoRA}: LoRA applied to different components in the pre-trained FEDformer: all decoder layers (\textit{FED-LoRA-Decoder}), the Frequency-Enhanced-Attention block (\textit{FED-LoRA-FEA}), or the Encoder, Decoder, and FEA (\textit{FED-LoRA-ALL}).
\item \textit{Transformer/Informer/Autoformer-LoRA-Encoder}: LoRA applied to the Encoder in baseline models: Transformer~\cite{vaswani2017transformer}, Informer~\cite{zhou2021informer}, and Autoformer~\cite{chen2021autoformer}.
\end{itemize}


\begin{figure*}
    \centering
    \scalebox{0.65}{
    \begin{subfigure}[b]{0.5\linewidth}
        \includegraphics[width=\linewidth]{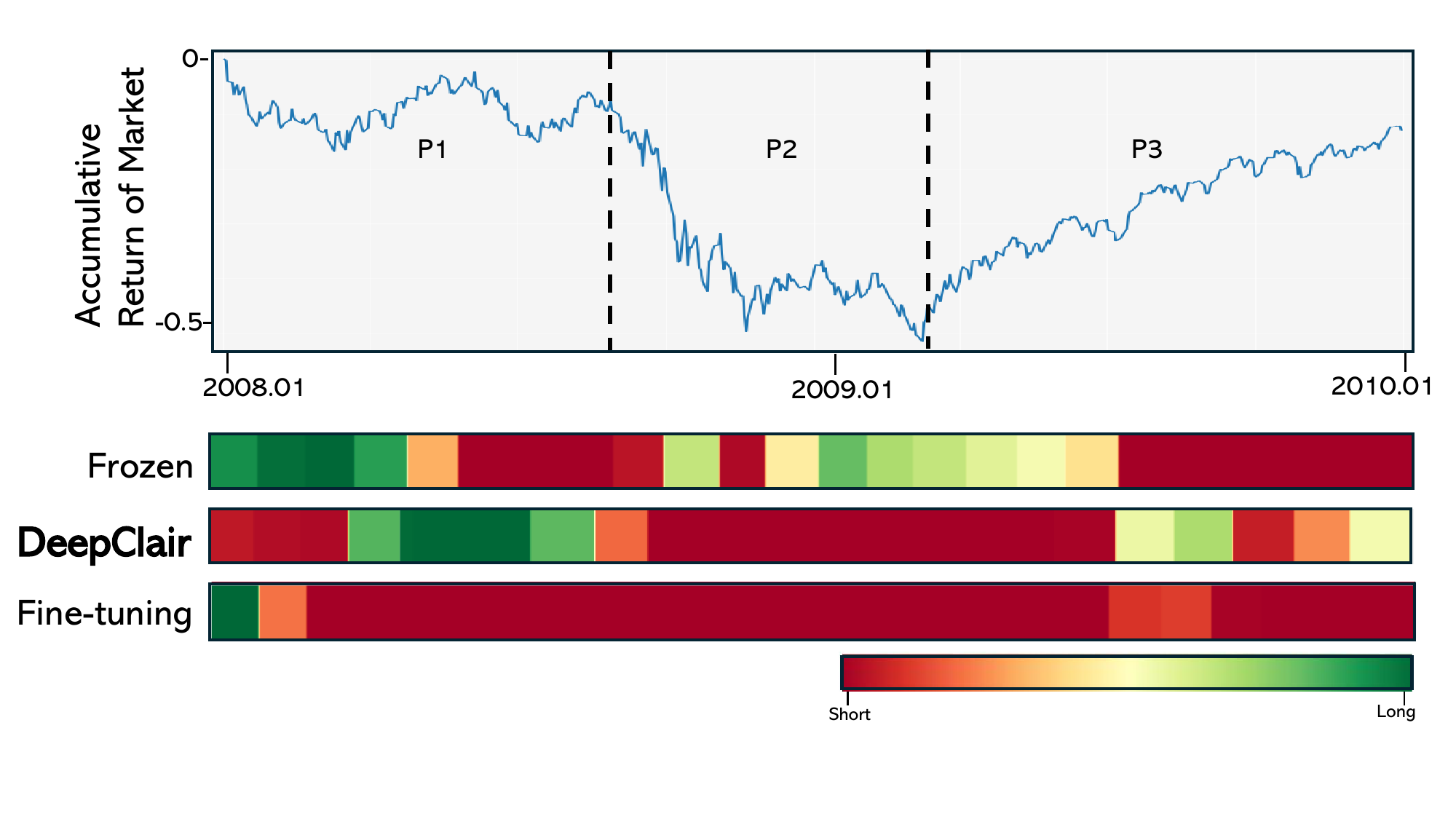}
        \caption{2008.01$\sim$2010.01, encompassing the 2008 Global Financial Crisis and subsequent recovery period.
        }
        \label{fig:results_main_rmse}
    \end{subfigure}
    \hspace{0.5em}%
    \begin{subfigure}[b]{0.5\linewidth}
        \includegraphics[width=\linewidth]{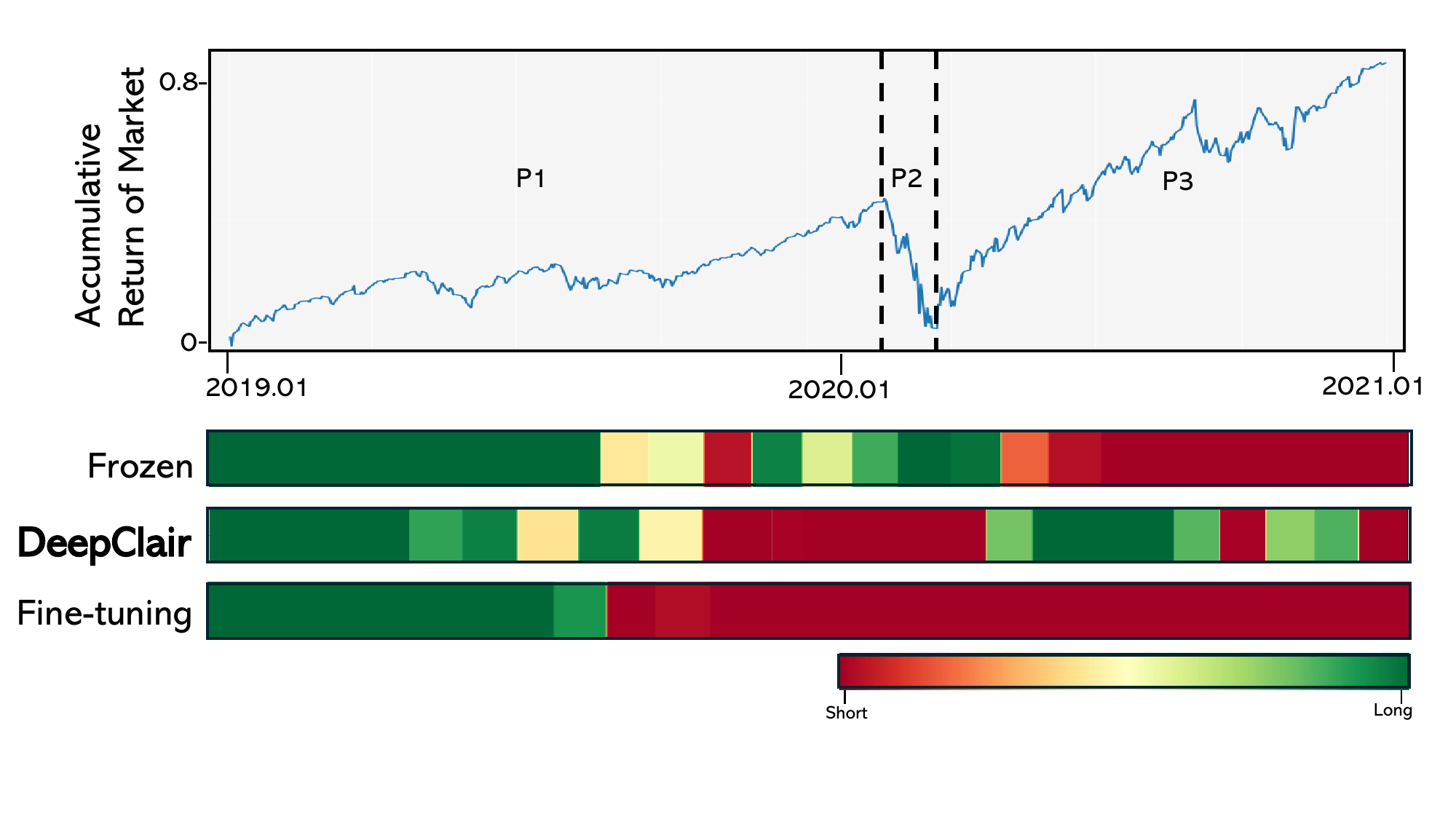}
        \caption{2019.01$\sim$2021.01, covering the market crash during the COVID-19 shock and the subsequent bullish period.
        }
        \label{fig:results_main_pcorr}
    \end{subfigure}
    }
    \caption{Nasdaq Index for two notable characteristics of market periods and the changing patterns of $\rho$.
    \textsc{DeepClair} adjusts its position during the bearish market of the crisis periods, transitioning from P1 to P2, and during the subsequent recovery period, transitioning from P2 to P3 in both (a) and (b).
    }
    \label{fig:qualitative_analysis}
\end{figure*}

As shown in Table~\ref{table:sub}, experimental results for ablations on optimization of the FEDformer used as \textsc{DeepClair}'s market forecasting module implicate several findings. 
We concluded that unfreezing all FEDformer parameters is not a feasible choice in optimizing \textsc{DeepClair} on the portfolio selection task.

In addition, experimental results for ablations on LoRA applied to FEDformer demonstrate that injecting the trainable matrices into the pre-trained FEDformer's encoder module yields the best investment performance in both datasets~\cite{hu2021lora}.
Since conventional transformers apply LoRA to its encoder and\slash or decoder module, we explored different options that also consider the FEDformer's FEA block.
Among different options regarding which sub-component of FEDformer to apply LoRA, \textit{FED-LoRA-FEA} yielded the worst performance along with \textit{FED-LoRA-All}.
We speculate that altering the pre-trained parameters of the FEA block rather disrupts \textsc{DeepClair}'s learning trajectory which should be aimed towards optimizing its portfolio selection ability.
Overall, applying LoRA to only the FEDformer's encoder module proved to be the optimal choice.


\subsection{RQ4. \textsc{DeepClair} exhibits adaptive investment behavior in drastically changing situations.}

In Figure~\ref{fig:qualitative_analysis}, we scrutinize \textsc{DeepClair}'s portfolio selection behavior by monitoring the parameter $\rho$, determining the ratio between long and short positions, and compare it with other design choices such as Frozen and Fine-tuning.

The Frozen scenario exposes the model's incapacity to adhere to a long-range investment strategy, potentially due to the forecasting model's output being treated as a noisy false signal.
Conversely, in the Fine-tuning scenario, the model tends to invest overwhelmingly in a coherent manner, suggesting a potential loss of forecasting ability and convergence toward a generic RL approach.

In our case, \textsc{DeepClair} demonstrates adaptive capabilities, effectively responding to both bullish and bearish market conditions. We posit that this adaptability is facilitated by LoRA, serving as a bridge to align the generalizability of the forecasting output with the RL structure.

Overall, our experimental results demonstrate three findings: 1) the significance of exploiting market forecasting in portfolio selection tasks, 2) the rationale behind utilizing FEDformer as market forecast module in \textsc{DeepClair}, and 3) the benefits of applying LoRA to FEDformer's encoder module when training \textsc{DeepClair} on portfolio selection task.



\section{Conclusion}
In this work, we introduced \textsc{DeepClair}, a novel RL portfolio selection system that leverages a market-forecasting module based on a frequency-enhanced transformer architecture pre-trained with supervised time-series forecasting.
Our approach involves two stages of training strategies: pre-training the forecasting module to establish forecasting abilities and fine-tuning the transferred module to determine \textsc{DeepClair}'s long-short investment positions through a portfolio selection RL training procedure. 
The portfolio selection task itself aims at formulating practical investment practices, making our work focused on creating strategies that can be effectively applied in real-world scenarios. 
In our experiments, we showcased our method's adaptive investments while demonstrating better returns.

%
Since \textsc{DeepClair} relies on deep learning architecture, capturing the causal relationship between inputs and investment decision-making can be challenging. 
However, we plan to conduct a thorough investigation into the interpretability of the backbone transformer attention architecture we use. Additionally, although our work follows the mainstream portfolio selection experiment settings~\cite{wang2021deeptrader,kim2023hadaps,lee2021maps}, we aim to explore more realistic and robust experimental settings. This includes accounting for transaction fees, incorporating other investment sources such as gold, cash, and cryptocurrencies, and experimenting with different date splits (train, test, validation) for investments.

This work was conducted in collaboration with investment experts at Shinhan Bank, aiming to develop practically usable advanced automated investment strategies using artificial intelligence.




\begin{acks}
This work was supported in part by the National Research Foundation of Korea (NRF-2023R1A2C3004176) and the MSIT (Ministry of Science and ICT), Korea, under the ICT Creative Consilience program (RS-2020-ll201819) supervised by the IITP (Institute for Information \& communications Technology Planning \& Evaluation).

Donghee Choi is additionally supported by the Horizon Europe project CoDiet. The CoDiet project is funded by the European Union under Horizon Europe grant number 101084642 and UK Research and Innovation (UKRI) under the UK government's Horizon Europe funding guarantee.
\end{acks}

\bibliographystyle{ACM-Reference-Format}
\balance
\bibliography{ref}

\appendix



\end{document}